\documentclass{article}

\usepackage{graphicx}
\usepackage{psfig}
\usepackage{epsfig}
\usepackage[round]{natbib}

\title{Probabilistic temperature forecasting: a summary of our recent research results}

\author{Stephen Jewson\footnote{\emph{Correspondence address}: RMS, 10 Eastcheap,
London, EC3M 1AJ, UK. Email: \texttt{x@stephenjewson.com}}}
\begin{document}

\maketitle

\begin{abstract}
We summarise the main results from a number of our recent articles on the subject of
probabilistic temperature forecasting.
\end{abstract}

\section{Introduction}

Industries such as finance, insurance, reinsurance and energy
have been using quantitative probabilistic
methods for many years. But when
these industries incorporate weather forecasts into their calculations
they tend to only use forecasts of the expectation of future
outcomes. Why don't they use probabilistic forecasts?
There are a number of reasons, including the following:

\begin{itemize}

\item Adequate probabilistic meteorological forecasts
are not available commercially.
There are two parts to this problem. First, that very few
forecast vendors produce probabilistic forecasts at all
and second that those that are produced are not correctly
calibrated.

\item There is considerable confusion in the academic
meteorological literature about what a probabilistic forecast even
is, with the words `ensemble forecast' and `probabilistic
forecast' often being used as if they are the same thing
\footnote{Our rather straightforward view is that a probabilistic forecast is a forecast
that gives probabilities, while an ensemble forecasts is a
forecast that consists of many members. They are not the same
thing: an ensemble forecast is not a probabilistic forecast
because it doesn't give probabilities, and a probabilistic forecast
is not an ensemble forecast because it doesn't contain members.
The two are, of course, related:
probabilistic forecasts can be made from single or ensemble forecasts using
statistical methods, and
probabilistic forecasts can be turned into ensemble forecasts by sampling.
From the point of view
of temperature forecasting, the main reason for running ensemble
forecasts is \emph{not} to generate probabilities, but to improve the
forecast of the expectation: see~\citet{jewson03i}.}.

\item There is considerable confusion about how ensemble forecasts, and in
particular the ensemble spread, should be interpreted and used.

\item The methods suggested in the academic literature for making
probabilistic forecasts are complicated, ad-hoc, poorly understood
and mostly contain obvious flaws.
Complex new methods have been introduced but have not been compared
with simpler methods.

\item Meteorologists have generally evaluated probabilistic
forecasts using methods they have invented themselves rather than
using well known and well understood methods from statistics. This
is not to say that the methods used by meteorologists are not
intrinsically good ones (see further comments on this below), but just
that they are not well understood or accepted by anyone outside
the rather narrow field of meteorological forecast verification.
There are methods, however, that are used throughout statistics to
solve essentially the same problem and that are well understood
by much of the statistical and scientific community.

\item Past forecasts are generally very hard to get hold of. Users
of forecasts are generally rather suspicious of the claims of
forecast vendors, and the fact that neither forecast vendors nor
modellers make their past forecasts available compounds this
problem. It also makes it hard for users to evaluate whether forecasts
are any good. We ourselves have extreme difficulties in obtaining
past forecasts, even for research purposes.

\end{itemize}

We would like to use probabilistic forecasts of site-specific
temperatures, and, given the problems listed above, have decided to
investigate ourselves how such forecasts might be produced.
Our hope is that the methods we have developed might be taken up
by forecast vendors from whom we could then obtain the resulting forecasts.
We
have published a series of articles describing the results of our
research, and the purpose of this current article is to summarise
the results thus far. The articles on which this summary is based
are
\citet{jewsonbz03a},
\citet{jewson03g},
\citet{jewson03i},
\citet{jewson03h},
\citet{jewson03aa},
\citet{jewson04c}.
There articles are all freely available on the internet, and comments are welcome.

\section{Modelling philosophy}

We believe that it is crucially important to use an appropriately
scientific modelling philosophy, and that such a philosophy is
generally lacking in the academic literature on this subject.
Our philosophy is as follows:

\begin{itemize}

\item To start with extremely simple models and build up, step by
step, to more complex models. At each stage we compare the more
complex models with the simpler models, and attempt to understand
what each new effect brings to the forecast.
As we will see below, this approach has led to significant insight
into what is important and what not in the generation of probabilistic forecasts.
We don't have any
particular axe to grind with respect to different techniques such
as classical statistics, Bayesian statistics or analogue methods.
However we do believe that classical statistics offers the
simplest models and
so we start there. If other methods can be shown to be materially
better than the classical statistical methods that we propose we
would be happy to adopt them.

\item To test everything empirically. For instance, it is often
claimed that the ensemble spread is a measure of the uncertainty
in a forecast, and this is assumed to be true by a number of the calibration
methods that have been suggested in the academic literature and
are used to make commercial probabilistic forecasts.
Whether spread really is a useful measure of the uncertainty in a
forecast may or may not be true and will likely depend on the forecast
model, the variable being predicted, and other factors. But this question
should \emph{always} be tested by including and excluding the
spread and observing the effect on the final forecast skill.
Considering the spread-skill correlation is not relevant to this question
since it does not consider the affects of spread on the skill of the
final forecast produced.

\item To try and find the optimum blend between information from
forecast models and from past forecast error statistics. There is
a general principle that should be followed when designing
calibration models:
forecast models tell us about variability, while only past
forecast error statistics can tell us constants. This is because
the processes that determine constants are not included in the
forecast models. This principle applies equally well to the mean
temperature, the forecast uncertainty and the forecast
correlation. For instance the mean level of uncertainty and
the amplitude of variability of the uncertainty must both come
from past forecast error statistics and should not be taken directly
from model output.

\end{itemize}

\section{Statement of the problem}

Weather forecast information is produced
from numerical models of the atmosphere.
The information from
these models, however, should not be thought of as consisting of a
prediction, but rather as a set of predictors.
These predictors consist of ensemble members, ensemble means and
single integrations, possibly from several models.
As mentioned above, it is dangerous to make
assumptions about what information there may be in any of these
predictors: everything should be tested empirically.

An additional source of information is contained in the
performance of forecasts in the past.

The challenge is to take these sources of information and
combine them to produce an optimal probabilistic prediction of the
weather.

\section{Forecast scoring}

Scoring probabilistic forecasts is all about putting samples from
the distribution of weather variability up against the
predicted distribution and comparing them. The generic statistical
question of comparing observations with a distribution was
addressed in~\citet{fisher1912} and~\citet{fisher1922}.
In these papers Fisher suggested the likelihood,
defined as the probability of the observations given the model,
as a useful measure. The likelihood rapidly gained wide acceptance
in statistics as the most appropriate way to measure the goodness
of fit of a distribution and has been used ever since.
There are many variations of the likelihood such as the various information
criteria (AIC, BIC, SIC). We use the likelihood (or
versions thereof) for all of our forecast scoring.

Academic meteorologists have tended to use other measures,
however, and in particular one measure known as the Brier score~\citep{brier}.
The Brier score is popular principally because it can be decomposed in a way
that is similar to a well known
decomposition of the mean square error.
We believe, however, that the Brier score is not a sensible generic measure for probabilistic forecasts.
Consider the following example: forecast 1 predicts a probability of 10\% for a certain event,
while forecast 2 predicts a probability of 0\%. The real probability is 4\%.
The Brier score chooses forecast 2 as best,
while we believe that the correct choice would be forecast 1.
This problem is also shared by derivatives of the Brier score such as the CRPS,
and is discussed in more detail in~\citet{jewson04a}.

\section{Forecast comparison}

Forecasts can be compared 3 ways: in sample, out of sample or in
real forecast mode. The latter is the ideal, but is seldom practical. In
sample testing can be used for the comparison of parametric calibration models.
If the models being compared have the same number of parameters then
likelihood scores can be used directly. If they do not then one has
to use one of the various information criteria to correct for over-fitting.
Out of sample testing should be used for all non-parametric
models. The results we describe below were based on a combination of
in and out of sample testing.

\section{Data}

All of our calibration experiments have been based on a single
year of daily ECMWF forecasts for London Heathrow. These have been
compared with the appropriate climate observations.
This data is deseasonalised, and calibration is applied to the anomalies.
We are well aware that using only a single year of data is not ideal. However
forecast models are changed rather frequently, and if much more
than a year of data is used it is unlikely to be stationary, thus
invalidating the whole idea of building a statistical model between
model output and observations.

\section{Calibration methods and results}

The calibration methods we have tested are listed below, along with a brief
description of the main results. The details are given in the relevant articles.
We fit all the models described below by maximising the likelihood: in some
cases this requires numerical methods.

\subsection{Linear regression}

Our first model is simply linear regression between temperature on day $i$ ($T_i$) and the ensemble
mean on day $i$ ($m_i$), which we write as

\begin{equation}
  T_i \sim N(\alpha+\beta m_i, \gamma)
\end{equation}

This model corrects biases using $\alpha$, optimally "damps" the variability of the ensemble mean
and merges optimally with climatology using $\beta$,
and predicts flow-independent uncertainty using $\gamma$.
The bias and the uncertainty produced by this model vary seasonally because of the deseasonalisation
and reseasonalisation steps.

Linear regression has been used for calibration
of forecasts of the expected temperature since at least the early
1970s~\citep{leith} but it has not always been realised that it also yields a
probabilistic forecast. We believe that all probabilistic calibration
experiments should start with linear regression as a baseline for
comparison. It is very simple, very easy to use, very well
understood and gives good results.

\subsection{Linear regression with the uncertainty predicted using the ensemble spread}

It is often claimed that the ensemble spread
contains information about the uncertainty in a forecast and so it
makes sense to adapt linear regression so that the uncertainty is
given by the ensemble spread rather than being fitted using past
forecast error statistics.
This gives

\begin{equation}
  T_i \sim N(\alpha+\beta m_i, s_i)
\end{equation}

where $s_i$ is the ensemble spread on day $i$.

When we tested this model in~\citet{jewson03g} we found
that the resulting forecasts performed significantly worse than
linear regression, and even worse than \emph{climatology} at short lead
times. The explanation for this is that the uncalibrated ensemble
spread underestimates the uncertainty in the forecast, which has
been well known for a number of years.

\subsection{Linear regression with the uncertainty predicted as a multiple of the ensemble spread}
\label{multiple}

An obvious way to overcome the problem
with the previous model is to scale the ensemble spread.
This is then similar to the methods suggested by~\citet{roulstons03} and~\citet{mylne02a}.

This gives:

\begin{equation}
  T_i \sim N(\alpha+\beta m_i, \delta s_i)
\end{equation}

When we tested this model we found that the resulting forecasts were
worse than those from linear regression. The reason for
this is that a single scaling factor transformation of
the ensemble spread sets \emph{both} the mean level of the uncertainty \emph{and}
the amplitude of the variability of the uncertainty
with only one free parameter.
The mean level of uncertainty is underestimated in the ensemble, and $\delta$
is forced to be greater than one to correct for this. However, a $\delta$ greater
than one also inflates the \emph{variability} in the uncertainty.
In the forecast data we looked at the variability of the uncertainty actually needs to be
\emph{deflated} (see below).

\subsection{Linear regression extended to include non-normality using kernel densities}

It has been suggested that the non-linear dynamics of the atmosphere
might lead to non-normality in the distribution of future temperatures.
For this reason it seems sensible to relax the assumption of normality in
the linear regression model, and we used a flexible kernel
density model for this purpose. We call
this model \emph{kernel regression}~\citep{jewson03h}:

\begin{equation}
  T_i \sim K(\alpha+\beta m_i, \gamma, \lambda)
\end{equation}

where $\lambda$ is the bandwidth of the kernel density.

The results from the kernel regression model showed no improvement over the normal model,
however, indicating that any non-normality present in the ensemble does not contain
useful information. We don't know whether this is because the ensemble doesn't contain much
normality, or because it \emph{does} contain non-normality but that that non-normality is unrealistic.

We do not conclude that such kernel methods should not be used. They perform as well as the
normal distribution, and are a reasonable alternative for those who dislike imposing highly
parametric models.

\subsection{Linear regression with the uncertainty predicted using a linear function of ensemble spread}

We have seen that a
simple scaling of the ensemble spread did not improve on linear
regression because it tries to calibrate the mean and the
amplitude of the variability of the ensemble spread at the same time.
In this model we
calibrate them separately. We call this model \emph{spread regression}
(see~\citet{jewsonbz03a} and~\citet{jewson03g}).

\begin{equation}
  T_i \sim N(\alpha+\beta m_i, \gamma+\delta s_i)
\end{equation}

The parameters in this model were all significantly different from
their default values out to day 10, justifying their inclusion in the model.
That $\delta$ is significantly different from zero indicates that there
is some kind of spread-skill relationship.

We find that while the mean of the
uncertainty is best predicted by increasing the ensemble spread
(as we saw in section~\ref{multiple})
the variability of the uncertainty is best predicted by \emph{decreasing}
the amplitude of
the variability of the ensemble spread (i{.}e{.} $\delta$ is less than 1).
In fact we find that in
the calibrated forecast the variability of the uncertainty is
rather small (between 5\% and 20\% of the mean level):
the calibrated forecasts show more or less the same
level of uncertainty every day with only small variations.
That singular-vector forecasting systems such as that used at ECMWF
would tend overestimate the amplitude of the variability
in the uncertainty in this way is to be expected~\citep{jewsonaz04}.

We have learnt an important lesson from the spread regression model:
that calibration schemes should treat the mean and the amplitude of the variability of the uncertainty
separately. A simple thought experiment is useful to test whether a calibration
scheme satisfies this requirement. Imagine that the ensemble spread contains no
information whatsoever: does the calibration scheme ignore it and estimate uncertainty
based on past forecast error statistics alone? If not, the scheme is unlikely to
be as good as the spread regression model, and may not even be as good as
linear regression. Interestingly, most of the published calibration schemes
(such as~\citet{roulstons03}, \citet{mylne02a} and~\citet{raftery})
and all the the commercially calibrated ensemble or probabilistic forecasts we have seen
fail this test.
We therefore recommend that these schemes and forecasts should not
be used until they have been adapted to calibrate the spread in a more appropriate way.

The results from the spread regression model were the first of our results to beat the
linear regression model, although the difference was very small and could not
be detected as significant in out of sample tests. Why is the benefit from using the spread so small?
Presumably because the variations in the calibrated uncertainty forecast are not large
relative to the mean uncertainty and relative to the variability in the mean temperature.
Why, then, are the fluctuations
in the predicted uncertainty so small? This could be either because the ensemble
does a poor job in predicting the fluctuations in the uncertainty, or because
the fluctuations in uncertainty are inherently small.
We don't know how to distinguish between these two cases.

Another lesson we have learnt is that a statistically significant spread skill relationship
does not guarantee that the ensemble spread is a useful predictor: we found a significant
relationship but saw only a tiny benefit from actually using the spread as a predictor.

Comparing the benefit of using the ensemble mean with the benefit of using the ensemble
spread~\citep{jewson03i} we see that the ensemble mean is dramatically more useful. The spread is \emph{most}
useful at the shortest leads, but even then the ensemble mean is still much \emph{more} useful.

\subsection{Linear regression with the uncertainty predicted using a linear function of ensemble
spread and non-normality using kernel densities}

Our next model is a combination of two of the previous models: we extend the spread regression
model to include non-normality using the kernel density, to give the kernel spread
regression model.

\begin{equation}
  T_i \sim K(\alpha+\beta m_i, \gamma+\delta s_i, \lambda)
\end{equation}

As with the kernel regression model we find no benefit from including non-normality~\citep{jewson03h}.

\subsection{Linear regression with seasonally varying parameters}

Everything in meteorology varies seasonally, and there is no
particular reason to think that the optimum parameters
in our calibration models should be any different. We thus tested
a number of the preceeding models but with seasonally varying
parameters~\citep{jewson04c}.

The most complex of the models we tested was:
\begin{equation}
  T_i \sim N(\alpha_i+\beta_i m_i, \gamma_i+\delta_i s_i)
\end{equation}

where

\begin{eqnarray}
  \alpha_i&=&\alpha_0+\alpha_s \mbox{sin} \theta_i+\alpha_c \mbox{cos} \theta_i \\\nonumber
  \beta_i &=&\beta _0+\beta_s  \mbox{sin} \theta_i+\beta_c  \mbox{cos} \theta_i \\\nonumber
  \gamma_i&=&\gamma_0+\gamma_s \mbox{sin} \theta_i+\gamma_c \mbox{cos} \theta_i \\\nonumber
  \delta_i&=&\delta_0+\delta_s \mbox{sin} \theta_i+\delta_c \mbox{cos} \theta_i
\end{eqnarray}

We also tested a number of models with seasonality in only some of the parameters.

Adding seasonally varying parameters gave a
\emph{huge} improvement in our forecasts, including a big
improvement in the linear correlation between the forecast and observed
anomalies. This was mainly due to
the inclusion of seasonally varying bias correlation, although making the
other parameters vary seasonally helped too. Interestingly there was a synergistic effect
whereby making all the parameters vary seasonally at once gave a greater benefit
than the sum of the benefits from making them vary seasonally individually.

The importance of seasonally varying bias correction can be interpreted as being due
to discrepancies between the seasonal cycles in the observations and the model.

\subsection{Correlation calibration}

In~\citet{jewson03aa} we considered how best to predict the
correlation between different days of a forecast. Our
results showed that correlations predicted from past forecast
error statistics were more accurate than those predicted
directly from the ensemble, but that
an 80-20 mix of past forecast errors
statistics and correlations based on the ensemble gave better
predictions than either alone.

\section{Summary of results}

We summarise our main results as follows:

\begin{itemize}

\item The linear regression model is a very good starting point for
the comparison of different calibration methods.

\item We have found it very hard to see any more than a tiny benefit from using the
ensemble spread.

\item We found it essential to calibrate the spread using at least
two degrees of freedom: one for calibrating the mean level of the spread and the
other for calibrating the amplitude of the variability of the spread.
We consider all other calibration methods that we have seen in the literature or
in the commercial sector to be flawed because they confuse the calibration
of the mean and the variability of the spread.

\item We did not find any benefit from using
non-normal rather than normal distributions.

\item Extending the linear regression model to include seasonally
varying parameters was \emph{extremely} beneficial.

\end{itemize}

Putting all this together, we currently recommend the seasonal
parameter linear regression model as both the best method for producing
probabilistic forecasts and an appropriate baseline for
judging new calibration methods.

\subsection{Recommendations for forecast users}

Forecast users should be wary of using the forecast calibration methods described in the
academic literature, and wary of the probabilistic and
ensemble forecast products currently available from forecast
vendors. We are not convinced that any of these methods have been
well tested or do as well as the seasonal-parameter linear
regression model described above, and they are certainly much more complex and
liable to model error.

\subsection{Recommendations for forecast vendors}

We believe that forecast vendors need to take a simple,
transparent and pragmatic approach to producing probabilistic
forecasts. They should start by using the seasonal-parameter
linear regression model given above (or something similar) and progress beyond that
model only when more complex models have been \emph{empirically proven} to be better.

\subsection{Recommendations for numerical modellers}

It is slightly disappointing that we haven't been able to find
more use for the ensemble spread. One of the reasons for this is that the
information contained in the ensemble spread is subtle and hard to
calibrate. A great help in this respect would be longer series of
past forecasts from stationary models.

\subsection{Recommendations for calibration research}

The production of probabilistic meteorological forecasts is still in its infancy,
as this article has showed. There are many areas for future
research. This includes:

\begin{itemize}

\item Trying to beat the models described
above, perhaps using different transformations of the ensemble
spread, perhaps using Bayesian methods, perhaps by calibrating
full fields rather than anomalies, perhaps using high resolution
forecasts, etc.

\item Testing the models described above on different locations.

\item Testing the models described above on longer forecast records.

\item Applying the same straightforward calibration philosophy to wind and precipitation forecasts.

\end{itemize}

\section{Legal statement}

SJ was employed by RMS at the time that this article was written.

However, neither the research behind this article nor the writing
of this article were in the course of his employment, (where 'in
the course of their employment' is within the meaning of the
Copyright, Designs and Patents Act 1988, Section 11), nor were
they in the course of his normal duties, or in the course of
duties falling outside his normal duties but specifically assigned
to him (where 'in the course of his normal duties' and 'in the
course of duties falling outside his normal duties' are within the
meanings of the Patents Act 1977, Section 39). Furthermore the
article does not contain any proprietary information or trade
secrets of RMS. As a result, the author is the owner of all the
intellectual property rights (including, but not limited to,
copyright, moral rights, design rights and rights to inventions)
associated with and arising from this article. The author reserves
all these rights. No-one may reproduce, store or transmit, in any
form or by any means, any part of this article without the
author's prior written permission. The moral rights of the author
have been asserted.

The contents of this article reflect the author's personal
opinions at the point in time at which this article was submitted
for publication. However, by the very nature of ongoing research,
they do not necessarily reflect the author's current opinions. In
addition, they do not necessarily reflect the opinions of the
author's employer.

\bibliography{summary}

\end{document}